\documentclass[10pt,aps,pre,twocolumn,floatfix,showpacs,superscriptaddress,raggedbottom]{revtex4-1}

\usepackage{ucs}
\usepackage[utf8x]{inputenc}
\usepackage[english]{babel}
\usepackage[T1]{fontenc}

\usepackage{color}      

\usepackage{amssymb,amsmath,bm}
\usepackage{graphicx,latexsym}
\usepackage{ulem}

\newcommand{\ket}[1]{ |#1\rangle}

\newcommand{\brainnf}[3]{ \langle #1 | #2 | #3 \rangle}
\newcommand{\pbra}[1]{ (#1|}
\newcommand{\pket}[1]{ |#1)}

\newcommand{\pbrainnf}[3]{ ( #1 | #2 | #3 )}
\newcommand{\expo}[1]{e^{#1}}

\newcommand{\sfrac}[3][]{\left( \frac{#2}{#3} #1 \right)}

\newcommand{\alg}[1]{\left| #1\right|}
\newcommand{\slaufa}[1]{\left\{ #1 \right\}}
\newcommand{\svig}[1]{\left( #1 \right)}
\newcommand{\horn}[1]{\left[ #1 \right]}

\newcommand{\ntext}[1]{_{\textrm{#1}}}

\newcommand{\pa}[3][]{\frac{\partial^{#1} #2}{\partial #3}}

\begin{document}

\title{Symmetric excitation and de-excitation of a cavity QED system}

\author{Olafur Jonasson}
\affiliation{Science Institute, University of Iceland,
        Dunhaga 3, IS-107 Reykjavik, Iceland}

\author{Chi-Shung Tang}
\email{cstang@nuu.edu.tw}
 \affiliation{Department of Mechanical Engineering,
  National United University, 1, Lienda, Miaoli 36003, Taiwan}

\author{Hsi-Sheng Goan}
\email{goan@phys.ntu.edu.tw}
 \affiliation{Department of Physics and Center for Theoretical Sciences,
National Taiwan University, Taipei 10617, Taiwan}
 \affiliation{Center for Quantum Science and Engineering,
 National Taiwan University, Taipei 10617, Taiwan}

\author{Andrei Manolescu}
\affiliation{Reykjavik University, School of Science and
Engineering, Menntavegur 1, IS-101 Reykjavik, Iceland}

\author{Vidar Gudmundsson}
\email{vidar@hi.is}
 \affiliation{Science Institute, University of Iceland,
        Dunhaga 3, IS-107 Reykjavik, Iceland}

%

\begin{abstract}
We calculate the time evolution of a cavity-QED system subject to a time dependent sinusoidal drive.
The drive is modulated by an envelope function with the shape of a pulse. The system consists of electrons
embedded in a semiconductor nanostructure which is coupled to a single mode quantized electromagnetic field.
The electron-electron as well as photon-electron interaction is treated exactly using ``exact numerical
diagonalization'' and the time evolution is calculated by numerically solving the equation of motion for
the system's density matrix. We find that the drive causes symmetric excitation and de-excitation where
the system climbs up the Jaynes-Cummings ladder and descends back down symmetrically into its original state.
This effect persists even in the ultra-strong coupling regime where the Jaynes-Cummings model is invalid.
We investigate the robustness of this symmetric behavior with respect to the drive de-tuning and pulse duration.
\end{abstract}



\maketitle

%
%

\section{Introduction}
\label{sec:intro}
A quantum two level system (TLS) interacting with a single mode of a quantized electromagnetic field is a
central topic within the scope of circuit quantum electrodynamics (QED). Typically, the light-matter
interaction is weak enough to warrant the use of some version of the Jaynes-Cummings (JC) model \cite{jaynes1963}
to describe both static \cite{Wallraff2004,Fink2008,Kasprzak2010} and time dependent systems
\cite{Alsing1992,Bishop2009,Liberato2009,Schmidt2010}. However, recent progress in the field of circuit QED has enabled
the fabrication of ultra-small mode-volume cavities where the light-matter interaction strength can be a
considerable fraction of a cavity photon energy. In this regime, evidence of the breakdown of the JC-model
(with the rotating wave approximation) has been observed in superconducting \cite{Niemczyk2010} and semiconductor
systems \cite{Gunter2009,Anappara2009}.

In previous work, we have gone beyond the TLS approximation and solved the many-body Schrödinger equation
exactly for electrons embedded in a semiconductor nanostructure, subject to a single mode quantized EM field
and an external classical magnetic field \cite{Jonasson2012}. The electron-electron and photon-electron
interactions were treated exactly using exact numerical diagonalization (for details see Ref. \cite{Jonasson2012_2}).
We predicted the failure of the JC-model (including anti-resonance terms) at high coupling strengths for a static
and closed system. Here, we expand on that work by adding an explicit time dependence to the total Hamiltonian and
investigate its dynamical properties.

We investigate the effects of time dependent addition to the total Hamiltonian which does not depend on
the cavity photon creation or annihilation operators.  In the language of the JC-model, this means that we
are perturbing the atomic term of the total Hamiltonian as opposed to the cavity field or interaction term.
The time dependent term is a sinusoidal drive which is modulated by an envelope function which varies slowly
compared with other characteristic time scales of the system.

The paper is organized as follows. In Sec. \ref{sec:descr} we describe the static part of the
Hamiltonian which contains the geometry of the semiconductor nanostructure as well as the
electron-electron and photon-electron interaction. In Sec. \ref{sec:time} we add a time dependent
drive to the total Hamiltonian and introduce the time dependent observables we are interested in
and how we calculate them. Results and concluding remarks are presented in Secs. \ref{sec:results}
and \ref{sec:conclusions} respectively.

\section{Description of the static system}
\label{sec:descr}

\begin{figure}[ht]
  \centering
  \includegraphics[width=3.41in]{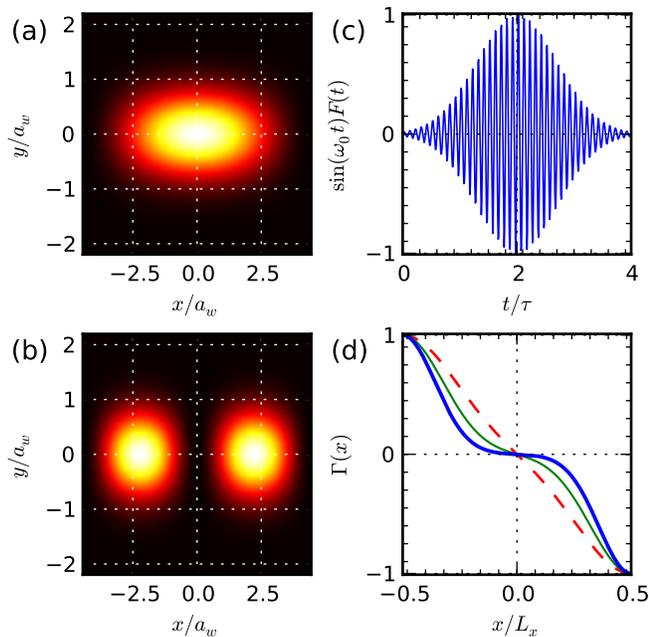}
  \caption{(Color online) Charge densities for the SES's $\ket{1}$ (a) and $\ket{2}$ (b) in arbitrary units
           for zero magnetic field. Dark is low density and bright is high. (c), sinusoidal pulse with a
           Gaussian envelope function with $\alpha=10$ which gives $10$ complete oscillations in the time
           interval $\tau$.  (d), spatial variation of the drive potential $\Gamma(x)$ for $\beta_xL_x=3$
           (dashed red line), $\beta_xL_x=4$ (thin green line) and $\beta_xL_x=5$ (thick blue line).
           $\Gamma$ is homogeneous in the $y$-direction so a 3D plot is not needed.}
  \label{fig:overview}
\end{figure}

We split the static part of the system's Hamiltonian $\mathcal H_0$ into four terms,
\begin{align}
 \label{eq:H0}
 \mathcal H_0 = \mathcal H_e + \mathcal H_{\textrm{EM}} + \mathcal H_p + \mathcal H_d \ .
\end{align}
They are the electronic part of the Hamiltonian ($\mathcal H_e$), the cavity field Hamiltonian
($\mathcal H_{\textrm{EM}}$), the paramagnetic ($\mathcal H_p$) and diamagnetic ($\mathcal H_d$)
electron-photon interaction terms. The electronic part can be written as
\begin{align}
 \label{eq:He}
 \mathcal H_e &= \mathcal H_w + \mathcal H_{\textrm{Coul}} \\
              &= \sum_i E_i d^\dagger_i d_i
              + \frac12\sum_{ijrs} \brainnf{ij}{V_{\textrm{Coul}}}{rs} d_i^\dagger d_j^\dagger d_s d_r\ ,
\end{align}
where $\mathcal H_w$ is the Hamiltonian of a finite quasi-one-dimensional
(Q1D) quantum wire with hard walls at $x=\pm L_x/2$ and a parabolic confinement in
the $y$-direction with characteristic
 energy $\hbar\Omega_0$, containing several non-interacting electrons.
The wire is subject to an external classical magnetic field $\mathbf
B = \nabla\times\mathbf A_{\textrm{ext}} = B\mathbf{\hat z}$. With the
magnetic field, the characteristic energy in the $y$-direction is modified
to $ \hbar\Omega_w=\hbar \sqrt{\Omega_0^2+\omega_c^2}$, where $\omega_c$
is the cyclotron frequency $\omega_c=qB/m^*$, with $q$ the positive elementary
charge and $m^*$ the electron effective mass. A natural length scale is
$a_w=\sqrt{\hbar/(m^*\Omega_w)}$. $d^\dagger_i$ and $d_i$ are fermionic
creation and annihilation operators of the single electron eigenstates
(SES) $\ket i$, with energies $E_i$, which include the effect of the
external magnetic field. 
Figs.\ \ref{fig:overview}(a) and \ref{fig:overview}(b) show the
charge density of the two lowest SES for zero magnetic field.

$\mathcal H_{\textrm{Coul}}$ contains the effect of the Coulomb interaction with the kernel
\begin{align}
 V_{\mathrm{Coul}}(\mathbf r,\mathbf r') = \frac{q^2/(4\pi\varepsilon)}{\sqrt{(x-x')^2+(y-y')^2+\eta^2}} \ ,
\end{align}
where $\eta$ is a small convergence parameter to regulate the singularity at $\mathbf r=\mathbf r'$.
We denote the many electron eigenstates of $\mathcal H_w$ as $\ket{\mu}$ (we use Latin indices for
single-electron states and Greek ones for the many-electron states) with the energy $E_\mu$. These
states are simply Slater determinants in the SES's $\ket i$. We denote eigenstates of $\mathcal H_e$
as $\pket\mu$ (denoted now with a rounded right bracket) with energy $\tilde E_\mu$. The states $\ket\mu$
and $\pket\mu$ are connected via a unitary transformation $\pket\mu=\mathcal V\ket\mu$. We calculate
$\mathcal V$ by diagonalizing $\mathcal H_e$ in the basis of the eigenstates of $\mathcal H_w$. Where needed, we use the notation $\pket{\mu}_{N}$ to
denote the $\mu$-th electronic state containing $N$ electrons. For example, $\pket{4}_2$ is the fourth
lowest two electron state. Note that we have a closed system so the number of electrons is a conserved quantity.

The cavity field EM Hamiltonian can be written as $\mathcal H_{\textrm{EM}}=\hbar\omega_p a^\dagger a$
where $a^\dagger$ and $a$ are bosonic creation and annihilation operators of a cavity photon with energy
$\hbar\omega_p$. The photon energy is typically chosen such that it matches a transition energy in the
quantum wire (i.e. the photon field is near a resonance).

By taking the long wavelength approximation for the quantized EM field, its vector potential can be written as
$\mathbf A_{\mathrm EM}=A_{\mathrm{EM}} \mathbf{\hat e}(a+a^\dagger)$ with $\mathbf{\hat e}$ a unit vector
pointing in the field's direction of polarization. The paramagnetic interaction term can then be written as
\begin{align}
 \label{eq:Hp}
 \mathcal H_p = \mathcal E_c \sum_{ij} d^\dagger_i d_j g_{ij}(a+a^\dagger)
\end{align}
where we have defined the electron-photon coupling strength $\mathcal E_c=qA_{\textrm{EM}}\Omega_w a_w$, which is
the characteristic energy scale for the photon-electron interaction. We have also introduced an effective dimensionless coupling tensor (DCT)
\begin{align}
 g_{ij}=\frac{a_w}{\hbar}\mathbf{\hat e}\cdot\brainnf{i}{\boldsymbol\pi}{j}\ ,
\end{align}
defining the coupling of individual single-electron states $\ket i$ and $\ket j$ by the photonic mode
where $\boldsymbol \pi=\mathbf p + q\mathbf A_{\textrm{ext}}$ is the mechanical momentum. It is useful
to generalize $g_{ij}$ to
\begin{align}
 \label{eq:Gmn}
 \mathcal G_{\mu\nu} = \sum_{i,j} g_{ij}\pbrainnf{\mu}{d_i^\dagger d_j}{\nu}
\end{align}
so that $\mathcal G_{\mu\nu}$ is the DCT which defines the coupling of many-electron states $\pket\mu$
and $\pket\nu$ by the photon field. From Eq.\ \eqref{eq:Gmn} we see that $\mathcal G_{\mu\nu}$
depends on the geometry of the quantum wire and on the magnetic field. Note that the action of $d_i$
and $d_i^\dagger$ is only known in the $\slaufa{\ket\mu}$ basis (of non-interacting Fock states).
To calculate for example $d_i\pket\mu$
we have to use $d_i\pket\mu = d_i\mathcal V \ket \mu$. This definition of $\mathcal{G}_{\mu\nu}$ makes
comparison with the JC-model easy, since the coupling energy $\mathcal E_{\textrm{JC}}$ in the JC model
(in the $\mathcal E_{\textrm{JC}}\sigma_x(a+a^\dagger)$ term) is related to the DCT via
$\mathcal E_{\textrm{JC}} = |\mathcal G_{\kappa\lambda}| \mathcal E_c$ where $\pket\kappa$ and
$\pket{\lambda}$ are the two states chosen for the TLS approximation (active states) \cite{Jonasson2012}.
We will refer to $\mathcal E_c$ as the electron-photon coupling strength (or simply the coupling strength)
and $\mathcal E_{\textrm{JC}}$ as the effective coupling strength.

The diamagnetic interaction term can be written as
\begin{align}
 \mathcal H_d = \frac{\mathcal E_c^2}{\hbar\Omega_w} \mathcal N_e \horn{\svig{a^\dagger a + \frac12}+\frac12\svig{a^\dagger a^\dagger + aa}} \ ,
\end{align}
where $\mathcal N_e$ is the electron number operator.

Now that we have defined all the terms in \eqref{eq:H0}, we can put the results together
and expand $\mathcal H_0$ in the $\slaufa{\pket{\mu}}$ basis,
\begin{align}
 \nonumber
 &\mathcal H_0 = \sum_\mu \tilde E_\mu \pket{\mu}\pbra{\mu} + \hbar\omega_p a^\dagger a 
 + \sum_{\mu\nu} \mathcal E_c \mathcal G_{\mu\nu} \pket{\mu}\pbra{\nu}(a+a^\dagger)\\
 &+ \frac{\mathcal E_c^2}{\hbar\Omega_w} N_e \slaufa{\svig{a^\dagger a + \frac12}+\frac12\svig{a^\dagger a^\dagger + aa}}
\end{align}
where $N_e$ is the number of electrons in the system.

Next we proceed to expand $\mathcal H_0$ in a basis containing a preset number of photons $\slaufa{\ket M}$
(eigenstates of $a^\dagger a$ with eigenvalue $M$). We have then expanded $\mathcal H_0$ in the complete
orthonormal basis $\slaufa{\ket{\breve \alpha}}=\slaufa{\pket\mu\otimes\ket M}$. From the diagonalization
process we obtain a unitary transformation $\mathcal U$ which satisfies
$\mathcal U \ket{\breve\alpha}= \pket{\breve \alpha}$ where $\pket{\breve \alpha}$ are eigenstates of
$\mathcal H_0$ satisfying $\mathcal H_0 \pket{\breve \alpha} = \breve E_\alpha \pket{\breve \alpha}$.
The states $\pket{\breve\alpha}$ are then used as a basis in time dependent calculations which are
covered in section \ref{sec:time}.

\section{Time dependent Hamiltonian}
\label{sec:time}
Now we add a time dependent drive to the Hamiltonian in the form (in first quantization)
\begin{align}
 W(t) = W_0 \Gamma(x,y)\sin(\omega_0 t)F(t) \ ,
\end{align}
where $W_0$ is the drive amplitude which has units of energy, $\Gamma(x,y)$ contains the
spatial dependence of $W(t)$ and $F(t)$ is an envelope function for the sinusoidal term
$\sin(\omega_0t)$. We choose $\omega_0$ such that $\hbar\omega_0$ matches some transition
energy in the quantum wire (i.e. the drive is near a resonance). We will refer to
$\hbar\omega_0$ ($\omega_0$) as the drive energy (frequency). In this work we will choose
$F(t)$ to be a pulse which varies slowly on the timescale $2\pi/\omega_0$ and satisfies
$F(\pm \infty)=0$ and $F(t)\leq1$. An example of such an envelope function is a Gaussian
\begin{align}
 \label{eq:gaussian}
 F(t)=\expo{-(t-t_0)^2/\tau^2}\ ,
\end{align}
where $\tau\gg2\pi/\omega_0$. It is useful to define the quantity $\alpha=\omega_0\tau/(2\pi)$
which gives the number of oscillations within the time interval $\tau$ (see Fig.\ \ref{fig:overview}(c)).
Note that a larger value of $\alpha$ translates into a longer pulse (not a faster oscillating pulse).
As for the geometric part we will use
\begin{align}
 \label{eq:gamma}
 \Gamma(x,y) = -\expo{-\beta_x^2(x-L_x/2)^2}+\expo{-\beta_x^2(x+L_x/2)^2} \ ,
\end{align}
so $\Gamma(x,y)=\Gamma(x)$ is a sum of two Gaussians of opposite sign that are centered on the opposite
end of the quantum wire (see Fig.\ \ref{fig:overview}(d)). The reasons for this choice of $\Gamma$ is
that it couples the time dependent part of the Hamiltonian strongly to a cavity field that is polarized
in the $x$-direction.

To add $W(t)$ to $\mathcal H_0$, we need to calculate the second quantization generalization of $W(t)$ in the $\slaufa{\pket{\mu}}$ basis using
\begin{align}
 \nonumber
 \mathcal W(t) = &\sum_{ij\mu\nu}\brainnf{i}{W}{j}\pbrainnf{\mu}{d_i^\dagger d_j}{\nu} \pket{\mu}\pbra{\nu} \\
 \label{eq:W}
               = & W_0 \sin(\omega_0t)F(t)\sum_{\mu\nu} \Gamma_{\mu\nu}\pket{\mu}\pbra{\nu}
\end{align}
where the matrix element $\Gamma_{\mu\nu}$ characterizes the coupling of individual Coulomb
interacting many-electron states $\pket\mu$ and $\pket\nu$ by $\mathcal W(t)$. In \eqref{eq:W},
$\mathcal W(t)$ is simply a unit operator in the photon Fock space since it does not depend on
$a$ or $a^\dagger$.

Now that we have the matrix representation of the total Hamiltonian
$\mathcal H = \mathcal H_0 + \mathcal W(t)$, we can calculate the time evolution of the system
by integrating the equation of motion
\begin{align}
 i\hbar \pa{\rho}{t} = \horn{\mathcal H,\rho} \ .
\end{align}
where $\rho$ is the density matrix of the system. This is done numerically using a Crank-Nicolson method.
We let the system start out in the ground state and investigate its excitation by $\mathcal W(t)$.

The observables we are interested in are the mean number of photons $\langle N \rangle(t)$ and energy
$E(t)=\langle \mathcal H \rangle(t)$ which can be calculated using
$\langle \mathcal A \rangle(t)=\textrm{Tr}\horn{\rho(t) \mathcal A(t)}$ where $\mathcal A$ is some observable.

\begin{figure}[top]
  \centering
  \includegraphics[width=3.41in]{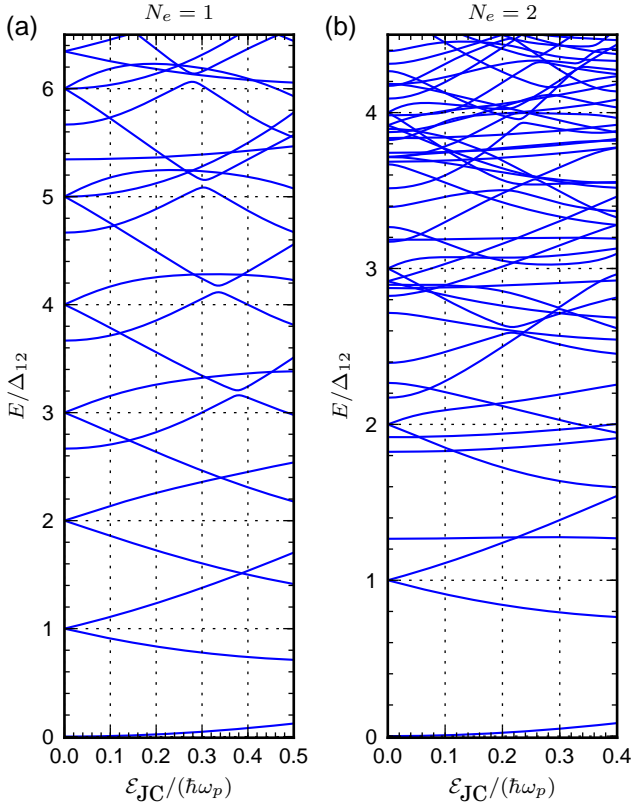}
  \caption{(Color online) (a), Energy spectra for one electron at zero external magnetic field as a functions
           of the effective coupling strength $\mathcal E_{\textrm{JC}}= |\mathcal G_{12}|\mathcal E_c$
           where $|\mathcal G_{12}|= 0.300$ and $\Delta_{12}=0.197$~meV. (b), same as in (a), but for two
           electrons with $|\mathcal G_{12}|= 0.676$ and $\Delta_{12}=0.521$~meV. Note that the energy
           spectrum has been shifted so that the ground state has zero energy at zero coupling strength.}
  \label{fig:spectrum}
\end{figure}

\section{Results}
\label{sec:results}

Throughout this section, we use $L_x=300$~nm, $\hbar\Omega_0=1.0$~meV, $\beta_xL_x=4$, $\alpha=90$,
$m^*=0.067m_e$ and $\varepsilon=12.4\varepsilon_0$ (GaAs parameters). Unless otherwise stated,
both the cavity photon energy $\hbar\omega_p$ as well as the drive energy $\hbar\omega_0$ are on
resonance between the electronic states $\pket1$ and $\pket2$ with de-tuning $\delta=10^{-4}\Delta_{12}$,
where $\Delta_{12}=\tilde E_2-\tilde E_1$ is the energy difference between the two lowest
electronic eigenstates. We let the system start out in its ground state and always use an
envelope function of the form in \eqref{eq:gaussian}. Where needed, we will differentiate
the de-tuning of the cavity field and the drive energy as $\delta_p$ and $\delta_d$ respectively.
We will only consider $x$-polarization for two reasons. First, it couples the states $\pket1$ and
$\pket2$ strongly (large $\alg{\mathcal G_{12}}$). Second, the quantum wire is approximately
(for small magnetic field) harmonic in the $y$-direction. We want to compare our results to
simpler TLS models and for a TLS approximation to be applicable, there needs to be some anharmonicity
present to minimize excitation to states outside of the TLS Hilbert space.

\begin{figure}[top]
  \centering
  \includegraphics[width=3.41in]{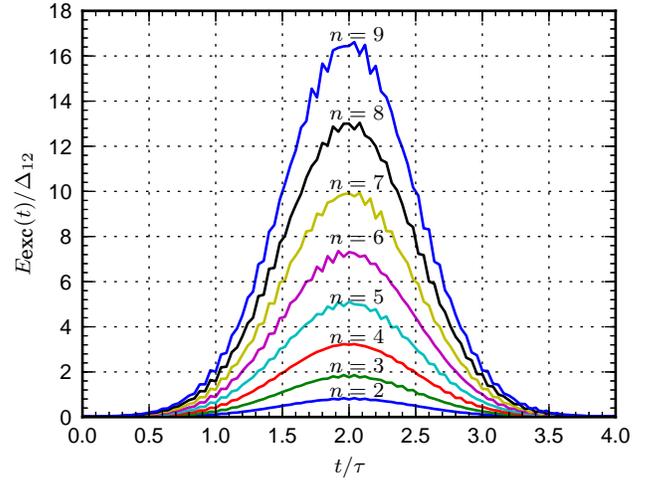}
  \caption{(Color online) Excitation energy vs time for a system containing one electron at zero magnetic field.
           The envelope function $F(t)$ is a Gaussian of the form in \eqref{eq:gaussian} with $t_0=2\tau$ and
           $\alpha=90$, giving $90$ oscillations in the time interval $\tau$.
           The effective coupling strength is $\mathcal E_{\textrm{JC}}/(\hbar\omega_p)\simeq0.13$.
           The system starts out in the ground state at $t=0$. The different curves are for different
           $W_0$ (see Eq.\ \eqref{eq:W}) where $W_0/\Delta_{12}=n\times0.428$.}
  \label{fig:Etime}
\end{figure}
\begin{figure}[ht]
  \centering
  \includegraphics[width=3.41in]{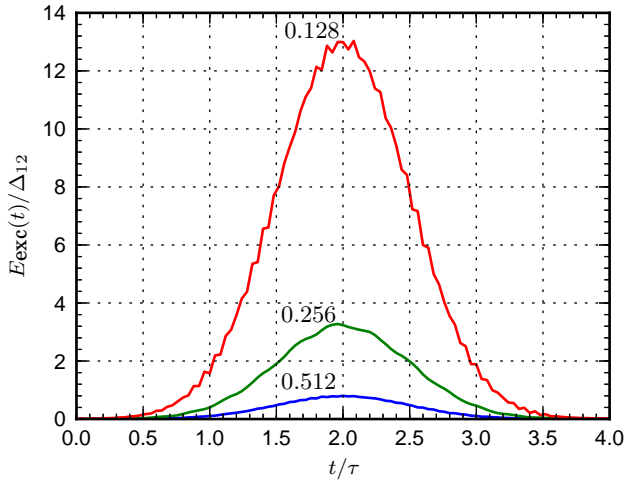}
  \caption{(Color online) Excitation energy vs time for a system containing one electron at zero magnetic field.
           All parameters are the same as in Fig.\ \ref{fig:Etime} except $\mathcal E\ntext{JC}$ is varied instead
           of $W_0$. Note that changing the electron-photon coupling shifts the ground state energy as can be seen
           in Fig.\ \ref{fig:spectrum}. That is why the above energy is shifted by the ground state energy $\breve E_0$.
           The three different numerical values $0.128$, $0.256$ and $0.512$ are the value of
           $\mathcal E\ntext{JC}/(\hbar\omega_p)$ used for for the corresponding curve.}
  \label{fig:Etime_Ec}
\end{figure}
\begin{figure}[ht]
  \centering
  \includegraphics[width=3.41in]{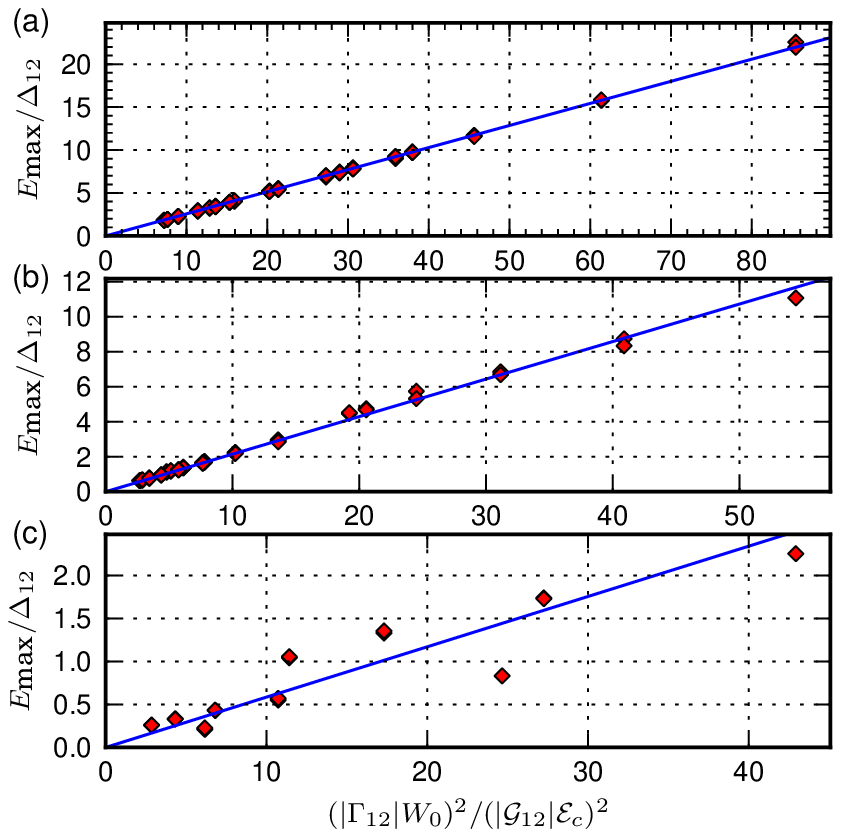}
  \caption{(Color online) Maximum excitation energy for many values of $\mathcal E_c$, $W_0$ and $B$ with
           $\mathcal E_c$ and $W_0$ in the range of that of Figs.\ \ref{fig:Etime} and \ref{fig:Etime_Ec}
           and $B$ in the range $0.0$-$0.5$~T. Results are shown for one (a), two (b) and three (c)
           electrons with $52$, $47$ and $18$ data points respectively (some data-points overlap and are
           not distinguishable on the graphs). The blue lines are least square fit with \eqref{eq:Eex}
           giving $A(1) = 0.26$, $A(2)=0.21$ and $A(3)=0.058$. The results are very robust for one and
           two electrons but start to show significant deviations for three electrons.}
  \label{fig:linfit}
\end{figure}

Fig.\ \ref{fig:spectrum} shows energy spectra for $\mathcal H_0$ with one and two electrons. From the figure we can see
$4$-$6$ of the lowest polariton pairs in the JC-ladder, along with other states that are not a part
of the TLS approximation. Note how much denser the two electron spectrum is. The range of coupling
strength is similar to what we use for time dependent calculations later on.

Fig.\ \ref{fig:Etime} shows the excitation energy $E\ntext{exc}(t)=E(t)-E(t=0)$ of the system as a
function of time for various driving field amplitudes ($W_0$). The most interesting aspect of the
figure is the symmetric excitation and de-excitation of the system. What is happening is that the
system is climbing up the Jaynes-Cummings ladder and descending back down symmetrically.
This would not be surprising for a low coupling strength where both the anti-resonant and
diamagnetic terms of the electron-photon interaction Hamiltonian can be ignored. In that case,
when both the drive and photon energy are on resonance, it is possible to find quasi stationary states
which are periodic in time (see Ref. \cite{Alsing1992}). The effect of the envelope function
is then to adiabatically tune the system's quasi ground state. The adiabatic nature of the envelope
function prevents transitions beyond this slowly shifting quasi ground state. When the envelope function
goes to zero again the system is adiabatically shifted to its original state. Like mentioned earlier,
this behavior is expected at low coupling strength, however,
for the plots in Fig.\ \ref{fig:Etime}, the electron-photon coupling is rather large
($\mathcal E_{\textrm{JC}}/(\hbar\omega_p)\simeq 0.13$). In fact, for this coupling strength,
a visible deviation from the JC-model is apparent in Fig.\ \ref{fig:spectrum}(a) where the
polariton splittings don't show the characteristic linear $\pm\sqrt{n}\mathcal E_{\textrm{JC}}$
splittings which we expect from the JC-model (for further comparison with the JC-model,
see Ref. \cite{Jonasson2012} or \cite{jonasson2012_thesis}). Another interesting aspect of
Fig.\ \ref{fig:Etime} is that the value of the maximum energy $E\ntext{max}$ at $t\simeq 2\tau$
scales quadratically with $W_0$. This will be addressed later in this work.

\begin{figure}[ht]
  \centering
  \includegraphics[width=3.41in]{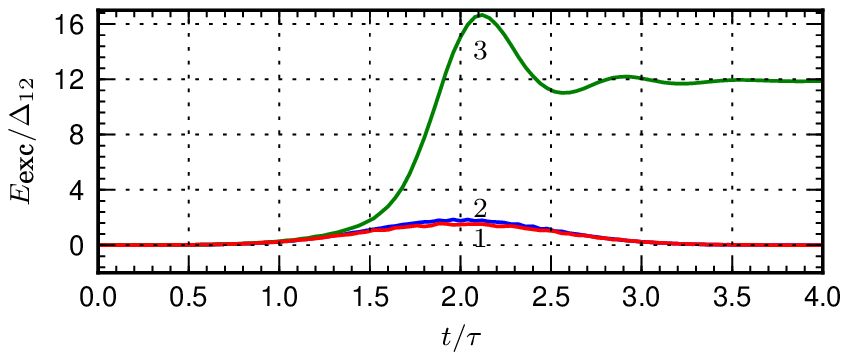}
  \caption{(Color online) Excitation energy vs time. All three plots are for one electron, $B=0$, $W_0=0.24$~meV,
           $\mathcal E_c=0.08$~meV and $\alg{\mathcal G_{12}}=0.3$, giving
           $\mathcal E_{\textrm{JC}}/(\hbar\omega_p)\simeq 0.13$. Curve $1$ shows the exact results using our full model.
           Curve $2$ shows results obtained using a TLS approximation using only two states for the electron
           Hilbert space basis where the diamagnetic interaction term is ignored. Curve $3$ shows results
           using a TLS approximation including the diamagnetic interaction term. Note how the TLS approximation
           with diamagnetic term drastically overestimates the excitation energy and how the system is not
           symmetrically de-excited.}
  \label{fig:m2_comparison}
\end{figure}

Fig.\ \ref{fig:Etime_Ec} shows the excitation energy of the system as a function of time for various
effective coupling strengths ($\mathcal E\ntext{JC}$), keeping the drive amplitude ($W_0$) constant
at $W_0/\Delta_{12}\simeq3.4$. Note that in $\mathcal E\ntext{JC}=\alg{\mathcal G_{12}}\mathcal E_c$,
only $\mathcal E_c$ can be varied since $\mathcal G_{\mu\nu}$ depends only on geometry.
The same excitation and de-excitation pattern is apparent, even for the ultrastrong coupling
$\mathcal E\ntext{JC}/(\hbar\omega_p)=0.512$. We did not perform calculations for higher coupling
because of issues with numerical convergence at higher coupling strengths (see Ref. \cite{Jonasson2012_2}
for detailed convergence calculations). Counter-intuitively, the maximum excitation energies at
$t\simeq2\tau$ in Fig.\ \ref{fig:Etime_Ec} is proportional to the inverse square of the coupling
strength $\mathcal E_c$. We can therefore see that the maximum excitation energy follows the
relationship $E\ntext{max}\propto (W_0/\mathcal E_c)^2$.

By doing calculations for many different values of $W_0$, $\mathcal E_c$ and magnetic field strength,
we were able to deduce the relationship
\begin{align}
 \label{eq:Eex}
 \frac{E\ntext{max}}{\Delta_{12}} = A(N_e) \sfrac{\alg{\Gamma_{12}}W_0}{\alg{\mathcal G_{12}}\mathcal E_c}^2 \ .
\end{align}
Where $A(N_e)$ is a dimensionless constant that depends only on the number of electrons
but not on the geometry of the system. Deducing the $(W_0/\mathcal E_c)^2$ relationship
was straightforward since $W_0$ and $\mathcal E_c$ are both parameters we can freely change.
However, to show the $|\Gamma_{12}/\mathcal G_{12}|^2$ relationship we had to vary the matrix
elements $\Gamma_{12}$ and $\mathcal G_{12}$. We did this by doing calculations for different
magnetic field strengths (see Fig.\ \ref{fig:linfit}). Varying the magnetic field changes the
geometry of the system and thereby changes both $\Gamma_{12}$ and $\mathcal G_{12}$. By doing
this we eliminate the possibility of the symmetric excitation de-excitation behavior being a
special case of the geometry we have chosen or due to the time reversal symmetry (which the magnetic field breaks).
Note that changing the magnetic field and electron
number alters the energy spectrum so $\Delta_{12}$ is not always the same. From Fig.\ \ref{fig:linfit}
we see that the agreement with Eq.\ \eqref{eq:Eex} is very good for one and two electrons while large
deviation is apparent for three electrons. The reason for this is that for more electrons,
the energy spectrum is much denser and previously inactive states start to have bigger influence.
Least square fit with \eqref{eq:Eex} gives $A(1) = 0.26$, $A(2)=0.21$ and $A(3)=0.058$ which shows
that a larger electron number translates into a weaker response to the drive.

\begin{figure}[t]
  \centering
  \includegraphics[width=3.41in]{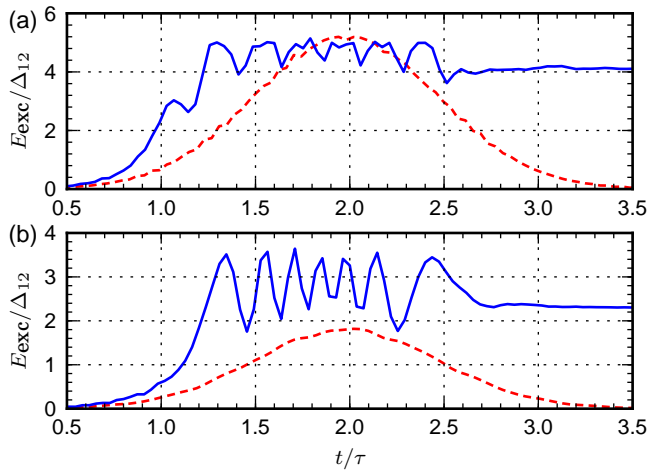}
  \caption{(Color online) (a), excitation energy vs time for two values of drive de-tuning for
           one electron with 
           $\mathcal E_{\textrm{JC}}/(\hbar\omega_p)\simeq0.13$, $W_0/\mathcal E_c=5$ and $B=0$.
           The two curves are for a de-tuning $\delta_d/\Delta_{12}=0.02$ (red-dashed) and
           $\delta_d/\Delta_{12}=0.05$ (solid blue), giving $\lambda\simeq 5.0$ and $\lambda=1.0$
           respectively. The symmetric behavior is lost in the latter case.
           (b), Same as in (a) except $\mathcal E_{\textrm{JC}}/(\hbar\omega_p)\simeq0.26$ and
           $W_0/\mathcal E_c=3$. The two curves are for a de-tuning $\delta_d/\Delta_{12}=0.01$
           (green-dashed) and $\delta_d/\Delta_{12}=0.1$ (solid black), giving $\lambda\simeq 14$
           and $\lambda \simeq 1.4$ respectively. The symmetric behavior is lost in the latter case.
           Note that in both (a) and (b), the cavity photon de-tuning is held constant at
           $\delta_p=10^{-4}\Delta_{12}$}
  \label{fig:detuning}
\end{figure}

We find that the mean number of photons follows the same relationship as the energy
\begin{align}
 \label{eq:Nmax}
 N\ntext{max} = A(N_e) \sfrac{\alg{\Gamma_{12}}W_0}{\alg{\mathcal G_{12}}\mathcal E_c}^2 \ ,
\end{align}
where $N\ntext{max}$ is the peak value of $\langle N \rangle(t)$. The constant of proportionality $A(N_e)$ is the same as for \eqref{eq:Eex}.

Because the geometric dependence of Eqs.\ \eqref{eq:Eex} and \eqref{eq:Nmax} is encoded in the
two matrix elements $\Gamma_{12}$ and $\mathcal G_{12}$ it is tempting to say that a TLS description
is applicable and we can ignore the inactive electronic states (all electronic states except
$\pket1$ and $\pket2$). This is true for small coupling strength where the JC-model can be used
and the diamagnetic interaction term is small. However, as can be seen from Fig.\ \ref{fig:m2_comparison},
a TLS description starts to fail at coupling strengths of $\mathcal E_{\textrm{JC}}/(\hbar\omega_p)\sim 0.1$.
What is surprising in Fig.\ \ref{fig:m2_comparison} is that for coupling strength
$\mathcal E_{\textrm{JC}}/(\hbar\omega_p)\sim 0.1$, including the diamagnetic interaction term in a TLS model
gives much less accurate results than if it is ignored. This behavior is observed when
$\mathcal E_{\textrm{JC}}/(\hbar\omega_p)\gtrsim 0.10$. It seems that the effects of the diamagnetic
interaction term are canceling the effects of the inactive states. However, for even higher coupling
strengths, the TLS model (with or without the diamagnetic interaction term) will fail because it can't
account for the complicated effect of inactive states in an energy spectrum such as the ones
in Fig.\ \ref{fig:spectrum}.

\begin{figure}[ht]
  \centering
  \includegraphics[width=3.41in]{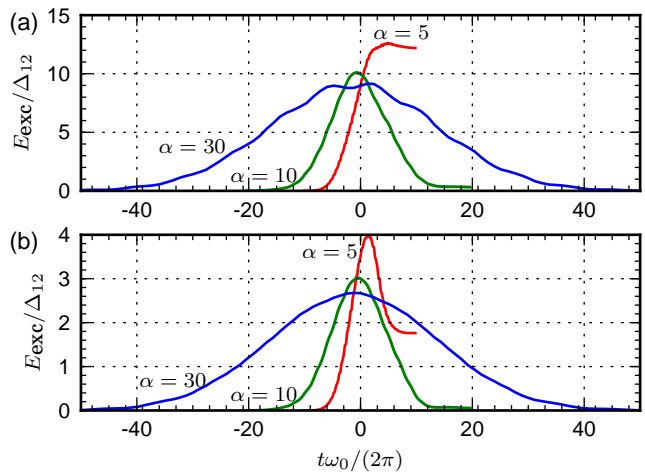}
  \caption{(Color online) Excitation energy vs time for different values of $\alpha$. (a), one electron,
           $B=0.2$~T, $\mathcal E_c=0.16$~meV, $W_0=0.48$~meV, $\Delta_{12}=0.172$~meV and
           $\mathcal E_{\textrm{JC}}/(\hbar\omega_p)=0.245$. (b), two electrons,
           $B=0.1$~T, $\mathcal E_c=0.08$~meV, $W_0=0.32$~meV, $\Delta_{12}=0.511$~meV and
           $\mathcal E_{\textrm{JC}}/(\hbar\omega_p)=0.101$. In both (a) and (b),
           we see that in order observe the symmetric behavior, we need to have
           $\alpha \gtrsim 10$. Note that the time axis has been shifted in both (a) and (b)
           such that the pulse maximum is at $t=0$.}
  \label{fig:alpha}
\end{figure}

We will end this section by investigating the robustness of the results. Fig.\ \ref{fig:detuning} shows
results with different values of the drive de-tuning. We find that symmetric excitation de-excitation
behavior persists as long as
\begin{align}
 \label{eq:lambda}
 \lambda \equiv  \frac{\Delta_{12}}{\delta_d} \sfrac{\alg{\mathcal G_{12}}\mathcal E_c}{\alg{\Gamma_{12}}W_0}^2\gg1 \ .
\end{align}
We also investigated what is the minimum pulse length needed to observe the
symmetric behavior (see Fig.\ \ref{fig:alpha}). We find that the minimum value
$\alpha\simeq10$ is needed. This means that we need at least $10$ complete oscillations
within the time period $\tau$. This is the same value of $\alpha$ which is used for
illustrative purposes in Fig.\ \ref{fig:overview}(c).

Throughout this work we used a Gaussian shaped envelope function. It is worth noting that the above mentioned symmetric behaviour
is not a special case for this choice. We tested this by varying the envelope function such that
$F(t)=\exp{\slaufa{-|t-t_0|^n/\tau^n}}$, with $n$ in the range $1$-$5$. This change does affect the shape of the excitation energy curves
such as those in Figs.\ \ref{fig:Etime} and \ref{fig:Etime_Ec} but the excitation and de-excition is still symmetric and has the same
maximum value as with a Gaussian shaped envelope function.

Finally, we note that so far we have only used envelope
functions that are symmetric around $t = t_0$. We found
that relaxing this restriction does change the symmetric
exciation and de-excitation behavior such as not being
symmetric about $t\simeq t_0$ anymore, but as long as the envelope function is slowly varying, the system still ends up in the
same state as it started in (the ground state).

\section{Concluding remarks}
\label{sec:conclusions}

We have described a method to compute the time evolution of a many-body, multi level,
Coulomb interacting electronic system which  is coupled to a single-mode quantized EM field.
The Hamiltonian contains explicit time dependence on the form of a sinusoidal drive which is
modulated by an envelope function. The envelope function is a Gaussian shaped pulse that varies
slowly compared to characteristic time scales of the system. Both the drive frequency and the
cavity photon frequency are on resonance with a transition in the electronic system. The only
approximations which are made are the finite size of single- and many-body bases as well as the
finite precision of the numerical integration. The convergence with respect to these parameters
is carefully controlled.

We observed a symmetric excitation and de-excitation of the system where the drive causes the
system to climb up the Jaynes-Cummings ladder. When the pulse dies out, the system descends down
the ladder back to its original (ground) state. This is not a simple adiabatic effect because even
though the envelope function varies slowly, the drive frequency matches that of the cavity photons
as well as a transition in the electronic nanostructure.

We found that when the system exhibits symmetric excitation and de-excitation, the maximum
expectation values of the energy and photon number follow a simple relationship which is
independent of the specific geometry, but does depend on the number of electrons in the system.
The effect is very robust for one electron, slightly less so for two electrons and considerable
deviation is observed for three electrons. This is due to the much denser many electron energy spectra,
where states which are not a part of the Jaynes-Cummings ladder start to have greater influence.

We compared the results of our full model to that of a TLS model. We did this by doing calculations
where only two states are used as a basis for the electronic part of the Hamiltonian. We found that
for the relatively weak coupling strength $\mathcal E_{\textrm{JC}}\simeq0.13$, including the
diamagnetic interaction term gives much worse results than if it is ignored. The results of the
TLS model without the diamagnetic term is in agreement with the results of the full model.

We tested robustness of the above mentioned symmetric behavior with respect to the magnitude
of the drive de-tuning. We found that the maximum allowed value of the relative de-tuning was $1$-$2$\%.
Note that this value depends strongly on geometry, coupling- and drive amplitude (see Eq.\ \eqref{eq:lambda}).
Finally, we investigated how long the pulse needs to be to see the symmetric effect. We found that
approximately $10$ complete oscillations are needed within the time period $\tau$, where $\tau$ characterizes
the duration of the pulse via Eq.\ \eqref{eq:gaussian}.\\

\begin{acknowledgments}
The authors acknowledge financial support from the Icelandic Research and Instruments
Funds, the Research Fund of the University of Iceland, the National Science Council of
Taiwan under contract No. NSC100-2112-M-239-001-MY3. HSG acknowledges support from the
National Science Council in Taiwan under Grant No. 100-2112-M-002-003-MY3, 
from the National Taiwan University under Grants No. 10R80911 and 10R80911-2, and
from the focus group program of the National Center for Theoretical Sciences, Taiwan.
\end{acknowledgments}


\begin{thebibliography}{14}%
\makeatletter
\providecommand \@ifxundefined [1]{%
 \@ifx{#1\undefined}
}%
\providecommand \@ifnum [1]{%
 \ifnum #1\expandafter \@firstoftwo
 \else \expandafter \@secondoftwo
 \fi
}%
\providecommand \@ifx [1]{%
 \ifx #1\expandafter \@firstoftwo
 \else \expandafter \@secondoftwo
 \fi
}%
\providecommand \natexlab [1]{#1}%
\providecommand \enquote  [1]{``#1''}%
\providecommand \bibnamefont  [1]{#1}%
\providecommand \bibfnamefont [1]{#1}%
\providecommand \citenamefont [1]{#1}%
\providecommand \href@noop [0]{\@secondoftwo}%
\providecommand \href [0]{\begingroup \@sanitize@url \@href}%
\providecommand \@href[1]{\@@startlink{#1}\@@href}%
\providecommand \@@href[1]{\endgroup#1\@@endlink}%
\providecommand \@sanitize@url [0]{\catcode `\\12\catcode `\$12\catcode
  `\&12\catcode `\#12\catcode `\^12\catcode `\_12\catcode `\%12\relax}%
\providecommand \@@startlink[1]{}%
\providecommand \@@endlink[0]{}%
\providecommand \url  [0]{\begingroup\@sanitize@url \@url }%
\providecommand \@url [1]{\endgroup\@href {#1}{\urlprefix }}%
\providecommand \urlprefix  [0]{URL }%
\providecommand \Eprint [0]{\href }%
\providecommand \doibase [0]{http://dx.doi.org/}%
\providecommand \selectlanguage [0]{\@gobble}%
\providecommand \bibinfo  [0]{\@secondoftwo}%
\providecommand \bibfield  [0]{\@secondoftwo}%
\providecommand \translation [1]{[#1]}%
\providecommand \BibitemOpen [0]{}%
\providecommand \bibitemStop [0]{}%
\providecommand \bibitemNoStop [0]{.\EOS\space}%
\providecommand \EOS [0]{\spacefactor3000\relax}%
\providecommand \BibitemShut  [1]{\csname bibitem#1\endcsname}%
\let\auto@bib@innerbib\@empty
\bibitem [{\citenamefont {Jaynes}\ and\ \citenamefont
  {Cummings}(1963)}]{jaynes1963}%
  \BibitemOpen
  \bibfield  {author} {\bibinfo {author} {\bibfnamefont {E.}~\bibnamefont
  {Jaynes}}\ and\ \bibinfo {author} {\bibfnamefont {F.}~\bibnamefont
  {Cummings}},\ }\href {\doibase 10.1109/PROC.1963.1664} {\bibfield  {journal}
  {\bibinfo  {journal} {Proceedings of the IEEE}\ }\textbf {\bibinfo {volume}
  {51}},\ \bibinfo {pages} {89 } (\bibinfo {year} {1963})}\BibitemShut
  {NoStop}%
\bibitem [{\citenamefont {Wallraff}\ \emph {et~al.}(2004)\citenamefont
  {Wallraff}, \citenamefont {Schuster}, \citenamefont {A.~Blais}, \citenamefont
  {Huang}, \citenamefont {Majer}, \citenamefont {Kumar}, \citenamefont
  {Girvin},\ and\ \citenamefont {Schoelkopf}}]{Wallraff2004}%
  \BibitemOpen
  \bibfield  {author} {\bibinfo {author} {\bibfnamefont {A.}~\bibnamefont
  {Wallraff}}, \bibinfo {author} {\bibfnamefont {D.~I.}\ \bibnamefont
  {Schuster}}, \bibinfo {author} {\bibfnamefont {L.~F.}\ \bibnamefont
  {A.~Blais}}, \bibinfo {author} {\bibfnamefont {R.-S.}\ \bibnamefont {Huang}},
  \bibinfo {author} {\bibfnamefont {J.}~\bibnamefont {Majer}}, \bibinfo
  {author} {\bibfnamefont {S.}~\bibnamefont {Kumar}}, \bibinfo {author}
  {\bibfnamefont {S.~M.}\ \bibnamefont {Girvin}}, \ and\ \bibinfo {author}
  {\bibfnamefont {R.~J.}\ \bibnamefont {Schoelkopf}},\ }\href {\doibase
  10.1038/nature02851} {\bibfield  {journal} {\bibinfo  {journal} {Nature}\
  }\textbf {\bibinfo {volume} {431}},\ \bibinfo {pages} {162} (\bibinfo {year}
  {2004})}\BibitemShut {NoStop}%
\bibitem [{\citenamefont {Fink}\ \emph {et~al.}(2008)\citenamefont {Fink},
  \citenamefont {G\"{o}ppl}, \citenamefont {Baur}, \citenamefont {Bianchetti},
  \citenamefont {Leek}, \citenamefont {Blais},\ and\ \citenamefont
  {Wallraff1}}]{Fink2008}%
  \BibitemOpen
  \bibfield  {author} {\bibinfo {author} {\bibfnamefont {J.~M.}\ \bibnamefont
  {Fink}}, \bibinfo {author} {\bibfnamefont {M.}~\bibnamefont {G\"{o}ppl}},
  \bibinfo {author} {\bibfnamefont {M.}~\bibnamefont {Baur}}, \bibinfo {author}
  {\bibfnamefont {R.}~\bibnamefont {Bianchetti}}, \bibinfo {author}
  {\bibfnamefont {P.~J.}\ \bibnamefont {Leek}}, \bibinfo {author}
  {\bibfnamefont {A.}~\bibnamefont {Blais}}, \ and\ \bibinfo {author}
  {\bibfnamefont {A.}~\bibnamefont {Wallraff1}},\ }\href {\doibase
  10.1038/nature07112} {\bibfield  {journal} {\bibinfo  {journal} {Nature}\
  }\textbf {\bibinfo {volume} {454}},\ \bibinfo {pages} {315} (\bibinfo {year}
  {2008})}\BibitemShut {NoStop}%
\bibitem [{\citenamefont {Kasprzak}\ \emph {et~al.}(2010)\citenamefont
  {Kasprzak}, \citenamefont {Reitzenstein}, \citenamefont {Muljarov},
  \citenamefont {Kistner}, \citenamefont {Schneider}, \citenamefont {Strauss},
  \citenamefont {H\"{o}fling}, \citenamefont {Forchel},\ and\ \citenamefont
  {Langbein}}]{Kasprzak2010}%
  \BibitemOpen
  \bibfield  {author} {\bibinfo {author} {\bibfnamefont {J.}~\bibnamefont
  {Kasprzak}}, \bibinfo {author} {\bibfnamefont {S.}~\bibnamefont
  {Reitzenstein}}, \bibinfo {author} {\bibfnamefont {E.~A.}\ \bibnamefont
  {Muljarov}}, \bibinfo {author} {\bibfnamefont {C.}~\bibnamefont {Kistner}},
  \bibinfo {author} {\bibfnamefont {C.}~\bibnamefont {Schneider}}, \bibinfo
  {author} {\bibfnamefont {M.}~\bibnamefont {Strauss}}, \bibinfo {author}
  {\bibfnamefont {S.}~\bibnamefont {H\"{o}fling}}, \bibinfo {author}
  {\bibfnamefont {A.}~\bibnamefont {Forchel}}, \ and\ \bibinfo {author}
  {\bibfnamefont {W.}~\bibnamefont {Langbein}},\ }\href {\doibase
  10.1038/nmat2717} {\bibfield  {journal} {\bibinfo  {journal} {Nature Mater}\
  }\textbf {\bibinfo {volume} {9}},\ \bibinfo {pages} {304} (\bibinfo {year}
  {2010})}\BibitemShut {NoStop}%
\bibitem [{\citenamefont {Alsing}\ \emph {et~al.}(1992)\citenamefont {Alsing},
  \citenamefont {Guo},\ and\ \citenamefont {Carmichael}}]{Alsing1992}%
  \BibitemOpen
  \bibfield  {author} {\bibinfo {author} {\bibfnamefont {P.}~\bibnamefont
  {Alsing}}, \bibinfo {author} {\bibfnamefont {D.-S.}\ \bibnamefont {Guo}}, \
  and\ \bibinfo {author} {\bibfnamefont {H.~J.}\ \bibnamefont {Carmichael}},\
  }\href {\doibase 10.1103/PhysRevA.45.5135} {\bibfield  {journal} {\bibinfo
  {journal} {Phys. Rev. A}\ }\textbf {\bibinfo {volume} {45}},\ \bibinfo
  {pages} {5135} (\bibinfo {year} {1992})}\BibitemShut {NoStop}%
\bibitem [{\citenamefont {Bishop}\ \emph {et~al.}(2008)\citenamefont {Bishop},
  \citenamefont {Chow}, \citenamefont {Koch}, \citenamefont {Houck},
  \citenamefont {Devoret}, \citenamefont {Thuneberg}, \citenamefont {Girvin},\
  and\ \citenamefont {Schoe}}]{Bishop2009}%
  \BibitemOpen
  \bibfield  {author} {\bibinfo {author} {\bibfnamefont {L.~S.}\ \bibnamefont
  {Bishop}}, \bibinfo {author} {\bibfnamefont {J.~M.}\ \bibnamefont {Chow}},
  \bibinfo {author} {\bibfnamefont {J.}~\bibnamefont {Koch}}, \bibinfo {author}
  {\bibfnamefont {A.~A.}\ \bibnamefont {Houck}}, \bibinfo {author}
  {\bibfnamefont {M.~H.}\ \bibnamefont {Devoret}}, \bibinfo {author}
  {\bibfnamefont {E.}~\bibnamefont {Thuneberg}}, \bibinfo {author}
  {\bibfnamefont {S.~M.}\ \bibnamefont {Girvin}}, \ and\ \bibinfo {author}
  {\bibfnamefont {R.~J.}\ \bibnamefont {Schoe}},\ }\href {\doibase
  doi:10.1038/nphys1154} {\bibfield  {journal} {\bibinfo  {journal} {Nature
  Physics}\ }\textbf {\bibinfo {volume} {5}},\ \bibinfo {pages} {105} (\bibinfo
  {year} {2008})}\BibitemShut {NoStop}%
\bibitem [{\citenamefont {De~Liberato}\ \emph {et~al.}(2009)\citenamefont
  {De~Liberato}, \citenamefont {Gerace}, \citenamefont {Carusotto},\ and\
  \citenamefont {Ciuti}}]{Liberato2009}%
  \BibitemOpen
  \bibfield  {author} {\bibinfo {author} {\bibfnamefont {S.}~\bibnamefont
  {De~Liberato}}, \bibinfo {author} {\bibfnamefont {D.}~\bibnamefont {Gerace}},
  \bibinfo {author} {\bibfnamefont {I.}~\bibnamefont {Carusotto}}, \ and\
  \bibinfo {author} {\bibfnamefont {C.}~\bibnamefont {Ciuti}},\ }\href
  {\doibase 10.1103/PhysRevA.80.053810} {\bibfield  {journal} {\bibinfo
  {journal} {Phys. Rev. A}\ }\textbf {\bibinfo {volume} {80}},\ \bibinfo
  {pages} {053810} (\bibinfo {year} {2009})}\BibitemShut {NoStop}%
\bibitem [{\citenamefont {Schmidt}\ \emph {et~al.}(2010)\citenamefont
  {Schmidt}, \citenamefont {Gerace}, \citenamefont {Houck}, \citenamefont
  {Blatter},\ and\ \citenamefont {T\"ureci}}]{Schmidt2010}%
  \BibitemOpen
  \bibfield  {author} {\bibinfo {author} {\bibfnamefont {S.}~\bibnamefont
  {Schmidt}}, \bibinfo {author} {\bibfnamefont {D.}~\bibnamefont {Gerace}},
  \bibinfo {author} {\bibfnamefont {A.~A.}\ \bibnamefont {Houck}}, \bibinfo
  {author} {\bibfnamefont {G.}~\bibnamefont {Blatter}}, \ and\ \bibinfo
  {author} {\bibfnamefont {H.~E.}\ \bibnamefont {T\"ureci}},\ }\href {\doibase
  10.1103/PhysRevB.82.100507} {\bibfield  {journal} {\bibinfo  {journal} {Phys.
  Rev. B}\ }\textbf {\bibinfo {volume} {82}},\ \bibinfo {pages} {100507}
  (\bibinfo {year} {2010})}\BibitemShut {NoStop}%
\bibitem [{\citenamefont {Niemczyk}\ \emph {et~al.}(2010)\citenamefont
  {Niemczyk}, \citenamefont {Deppe}, \citenamefont {Huebl}, \citenamefont
  {Menzel}, \citenamefont {Hocke}, \citenamefont {Schwarz}, \citenamefont
  {Garcia-Ripoll}, \citenamefont {Zueco}, \citenamefont {H\"{u}mmer},
  \citenamefont {Solano}, \citenamefont {Max},\ and\ \citenamefont
  {Gross}}]{Niemczyk2010}%
  \BibitemOpen
  \bibfield  {author} {\bibinfo {author} {\bibfnamefont {T.}~\bibnamefont
  {Niemczyk}}, \bibinfo {author} {\bibfnamefont {F.}~\bibnamefont {Deppe}},
  \bibinfo {author} {\bibfnamefont {H.}~\bibnamefont {Huebl}}, \bibinfo
  {author} {\bibfnamefont {E.~P.}\ \bibnamefont {Menzel}}, \bibinfo {author}
  {\bibfnamefont {F.}~\bibnamefont {Hocke}}, \bibinfo {author} {\bibfnamefont
  {M.~J.}\ \bibnamefont {Schwarz}}, \bibinfo {author} {\bibfnamefont {J.~J.}\
  \bibnamefont {Garcia-Ripoll}}, \bibinfo {author} {\bibfnamefont
  {D.}~\bibnamefont {Zueco}}, \bibinfo {author} {\bibfnamefont
  {T.}~\bibnamefont {H\"{u}mmer}}, \bibinfo {author} {\bibfnamefont
  {E.}~\bibnamefont {Solano}}, \bibinfo {author} {\bibfnamefont
  {A.}~\bibnamefont {Max}}, \ and\ \bibinfo {author} {\bibfnamefont
  {R.}~\bibnamefont {Gross}},\ }\href {\doibase 10.1038/nphys1730} {\bibfield
  {journal} {\bibinfo  {journal} {Nature Physics}\ }\textbf {\bibinfo {volume}
  {6}},\ \bibinfo {pages} {772–776} (\bibinfo {year} {2010})}\BibitemShut
  {NoStop}%
\bibitem [{\citenamefont {G\"{u}nter}\ \emph {et~al.}(2009)\citenamefont
  {G\"{u}nter}, \citenamefont {Anappara}, \citenamefont {Hees}, \citenamefont
  {Sell}, \citenamefont {Biasiol}, \citenamefont {Sorba}, \citenamefont
  {Liberato}, \citenamefont {Ciuti}, \citenamefont {Tredicucci}, \citenamefont
  {Leitenstorfer},\ and\ \citenamefont {Huber}}]{Gunter2009}%
  \BibitemOpen
  \bibfield  {author} {\bibinfo {author} {\bibfnamefont {G.}~\bibnamefont
  {G\"{u}nter}}, \bibinfo {author} {\bibfnamefont {A.~A.}\ \bibnamefont
  {Anappara}}, \bibinfo {author} {\bibfnamefont {J.}~\bibnamefont {Hees}},
  \bibinfo {author} {\bibfnamefont {A.}~\bibnamefont {Sell}}, \bibinfo {author}
  {\bibfnamefont {G.}~\bibnamefont {Biasiol}}, \bibinfo {author} {\bibfnamefont
  {L.}~\bibnamefont {Sorba}}, \bibinfo {author} {\bibfnamefont {S.~D.}\
  \bibnamefont {Liberato}}, \bibinfo {author} {\bibfnamefont {C.}~\bibnamefont
  {Ciuti}}, \bibinfo {author} {\bibfnamefont {A.}~\bibnamefont {Tredicucci}},
  \bibinfo {author} {\bibfnamefont {A.}~\bibnamefont {Leitenstorfer}}, \ and\
  \bibinfo {author} {\bibfnamefont {R.}~\bibnamefont {Huber}},\ }\href
  {\doibase 10.1038/nature07838} {\bibfield  {journal} {\bibinfo  {journal}
  {Nature}\ }\textbf {\bibinfo {volume} {458}},\ \bibinfo {pages} {178}
  (\bibinfo {year} {2009})}\BibitemShut {NoStop}%
\bibitem [{\citenamefont {Anappara}\ \emph {et~al.}(2009)\citenamefont
  {Anappara}, \citenamefont {De~Liberato}, \citenamefont {Tredicucci},
  \citenamefont {Ciuti}, \citenamefont {Biasiol}, \citenamefont {Sorba},\ and\
  \citenamefont {Beltram}}]{Anappara2009}%
  \BibitemOpen
  \bibfield  {author} {\bibinfo {author} {\bibfnamefont {A.~A.}\ \bibnamefont
  {Anappara}}, \bibinfo {author} {\bibfnamefont {S.}~\bibnamefont
  {De~Liberato}}, \bibinfo {author} {\bibfnamefont {A.}~\bibnamefont
  {Tredicucci}}, \bibinfo {author} {\bibfnamefont {C.}~\bibnamefont {Ciuti}},
  \bibinfo {author} {\bibfnamefont {G.}~\bibnamefont {Biasiol}}, \bibinfo
  {author} {\bibfnamefont {L.}~\bibnamefont {Sorba}}, \ and\ \bibinfo {author}
  {\bibfnamefont {F.}~\bibnamefont {Beltram}},\ }\href {\doibase
  10.1103/PhysRevB.79.201303} {\bibfield  {journal} {\bibinfo  {journal} {Phys.
  Rev. B}\ }\textbf {\bibinfo {volume} {79}},\ \bibinfo {pages} {201303}
  (\bibinfo {year} {2009})}\BibitemShut {NoStop}%
\bibitem [{\citenamefont {Jonasson}\ \emph
  {et~al.}(2012{\natexlab{a}})\citenamefont {Jonasson}, \citenamefont {Tang},
  \citenamefont {Goan}, \citenamefont {Manolescu},\ and\ \citenamefont
  {Gudmundsson}}]{Jonasson2012}%
  \BibitemOpen
  \bibfield  {author} {\bibinfo {author} {\bibfnamefont {O.}~\bibnamefont
  {Jonasson}}, \bibinfo {author} {\bibfnamefont {C.-S.}\ \bibnamefont {Tang}},
  \bibinfo {author} {\bibfnamefont {H.-S.}\ \bibnamefont {Goan}}, \bibinfo
  {author} {\bibfnamefont {A.}~\bibnamefont {Manolescu}}, \ and\ \bibinfo
  {author} {\bibfnamefont {V.}~\bibnamefont {Gudmundsson}},\ }\href {\doibase
  10.1088/1367-2630/14/1/013036} {\bibfield  {journal} {\bibinfo  {journal}
  {New Journal of Physics}\ }\textbf {\bibinfo {volume} {14}},\ \bibinfo
  {pages} {013036} (\bibinfo {year} {2012}{\natexlab{a}})}\BibitemShut
  {NoStop}%
\bibitem [{\citenamefont {Jonasson}\ \emph
  {et~al.}(2012{\natexlab{b}})\citenamefont {Jonasson}, \citenamefont {Tang},
  \citenamefont {Goan}, \citenamefont {Manolescu},\ and\ \citenamefont
  {Gudmundsson}}]{Jonasson2012_2}%
  \BibitemOpen
  \bibfield  {author} {\bibinfo {author} {\bibfnamefont {O.}~\bibnamefont
  {Jonasson}}, \bibinfo {author} {\bibfnamefont {C.-S.}\ \bibnamefont {Tang}},
  \bibinfo {author} {\bibfnamefont {H.-S.}\ \bibnamefont {Goan}}, \bibinfo
  {author} {\bibfnamefont {A.}~\bibnamefont {Manolescu}}, \ and\ \bibinfo
  {author} {\bibfnamefont {V.}~\bibnamefont {Gudmundsson}},\ }\href
  {http://arxiv.org/abs/1203.5980} {\enquote {\bibinfo {title} {Nonperturbative
  approach to circuit quantum electrodynamics},}\ }\bibinfo {howpublished}
  {arXiv:1203.5980} (\bibinfo {year} {2012}{\natexlab{b}})\BibitemShut
  {NoStop}%
\bibitem [{\citenamefont {Jonasson}(2012)}]{jonasson2012_thesis}%
  \BibitemOpen
  \bibfield  {author} {\bibinfo {author} {\bibfnamefont {O.}~\bibnamefont
  {Jonasson}},\ }\href {http://hdl.handle.net/1946/10771} {\enquote {\bibinfo
  {title} {Nonperturbative approach to circuit quantum electromagnetics},}\
  }\bibinfo {howpublished} {Master's Thesis, University of Iceland} (\bibinfo
  {year} {2012})\BibitemShut {NoStop}%
\end{thebibliography}

%

%
%
\end{document}